\documentclass[preprint,12pt]{elsarticle}

\usepackage{epsfig}
\usepackage{graphics,graphics} 
\usepackage{color}

\usepackage{amssymb,amsmath,amsthm}
\newtheorem{mydef}{Definition} 
\everymath{\displaystyle} 

\journal{Physica D}

\begin{document}

\begin{frontmatter}

\title{Causation Entropy Identifies Indirect Influences, \\ Dominance of Neighbors and Anticipatory Couplings}

\author{Jie~Sun}
\ead{sunj@clarkson.edu}
\author{Erik~M.~Bollt}
\ead{bolltem@clarkson.edu}
\address{Department of Mathematics, Clarkson University, Potsdam, NY 13699-5815}

\begin{abstract}
Inference of causality is central in nonlinear time series analysis and science in general. A popular approach to infer causality between two processes is to measure the information flow between them in terms of transfer entropy. Using dynamics of coupled oscillator networks, we show that although transfer entropy can successfully detect information flow in two processes, it often results in erroneous identification of network connections under the presence of indirect interactions, dominance of neighbors, or anticipatory couplings. {Such effects are found to be profound for time-dependent networks.}
To overcome these limitations, we develop a measure called {\it causation entropy}, and show that its application can lead to reliable identification of true couplings.
\end{abstract}


\end{frontmatter}

\section{Introduction}
\label{}
The long-standing puzzle of ``what causes what", formally known as the problem of causality inference, is challenging yet central in science~\cite{Hollanda1986,Pearl2009}. 
Understanding causal relationship between events has important implications in a wide range of areas including as examples social perception~\cite{Heider1944}, epidemiology~\cite{Rothman2005}, and econometrics~\cite{Heckman2008}.
It is the reliable inference of causality that allows one to untangle complex causal interactions, make predictions, and ultimately design intervention strategies.

Traditional approach of inferring causality between two stochastic processes is to perform the Granger causality test~\cite{Granger1969}. A main limitation of this test is that it can only provide information about {\it linear} dependence between two processes, and therefore fails to capture intrinsic nonlinearities that are common in real-world systems. 
To overcome this difficulty, Schreiber developed the concept of {\it transfer entropy} between two processes~\cite{Schreiber2000PRL}. Transfer entropy measures the uncertainty reduction in inferring the future state of a process by learning the (current and past) states of another process.
Being an asymmetric measure by design, transfer entropy is often used to infer the directionality of information flow and further the causality between two processes~\cite{Schindler2007,Vejmelka2008PRE}. Recently, it becomes increasingly popular to use transfer entropy for causality inference in networks of neurons~\cite{Honey2007PNAS,Vicente2011JCN} and in coupled dynamical systems with parameter mismatches~\cite{Bollt2012IJBC}, anticipatory couplings~\cite{Hahs2011PRL}, and time delays~\cite{Runge2012PRL}.
However, despite the overwhelming number of proposed applications, a clear interpretation of the resulting relationship inferred by transfer entropy is lacking.

In this paper, we study information transfer in the dynamics of small-scale coupled oscillators networks. We show by several examples that causal relationship inferred by transfer entropy are often misleading when the underlying system contains indirect connections, dominance of neighboring dynamics, or anticipatory couplings. To account for these effects, we develop a measure called {\it causation entropy (CSE)}, and show that its appropriate application reveals true coupling structures of the underlying dynamics.

\section{Information Theory and Dynamical Systems}
\label{}
In this section we introduce the mathematical tools used in this study, which include elements from both dynamical systems and information theory.

\subsection{Dynamical System as a Stochastic Process}
Our focus of this paper is on discrete dynamical systems of the form
\begin{equation}\label{eq:1}
x_{t+1} = f(x_t),
\end{equation}
where $x_t\in\mathcal{D}\subset\mathbb{R}^m$ is the state variable and $f:\mathcal{D}\rightarrow\mathcal{D}$ is the dynamic rule of the system.
A trajectory (or orbit) $\{x_t\}$ of Eq.~\eqref{eq:1} naturally represents a time series. 
For a continuous dynamical system $\dot{x}=f(x)$, a time series can be obtained by sampling its continuous trajectory at discrete time points. The time points are often chosen to spread uniformly in time or to be the times instances at which the trajectory intersects a given manifold that is transversal to the trajectory, called a {\it Poincar\'{e} section}~\cite{Smale2004}.

A natural bridge between dynamical systems and information theory is the formulation of symbolic dynamics, which requires discretization of the phase space. In particular, a finite {\it topological partition} $P=\{P_1,..,P_m\}$ of the phase space $\mathcal{D}$ is a collection of pairwise disjoint sets in $\mathcal{D}$ whose union is $\mathcal{D}$~\cite{Munkres2000}.
Defining the associated set of symbols $\Omega=\{1,2,\dots,m\}$, one can transform a trajectory $\{x_t\}$ into a 
{\it symbolic sequence} $\{s_t\}$, where $s_t$ is defined by~\cite{Lind1995,Kitchens1998}
\begin{equation}
	x_t\in P_i\subset\mathcal{D} \Rightarrow s_t = i\in\Omega.
\end{equation}

Viewing $\Omega$ as the sample space, the symbolic sequence $\{s_t\}$ can be seen as a time series of a stochastic process.
Define a probability measure over the partition $P$, as
\begin{equation}
	\mu: P \rightarrow \mathbb{R}.
\end{equation}
If $\mu$ is {\it invariant} under the dynamics, then~\cite{Eckmann1985RMP,Bollt2013}
\begin{equation}
	\mbox{Prob}(s_t=i) = \mu(i),~\mbox{~$\forall~i\in\Omega,~t\in\mathbb{R}$}.
\end{equation}
A partition $P$ is called a {\it Markov partition} if it gives rise to a stochastic process that is Markovian, i.e., future states of the process depends only on its current state, and not the past states~\cite{Cover2006,Bollt2005}.

\subsection{Information-Theoretical Measures: Entropy, Mutual Information and Transfer Entropy}
{Consider a discrete random variable $X$ whose probability mass function is denoted by $p(x)=\mbox{Prob}(X=x)$.
To quantify the unpredictability of $X$, one can calculate its {\it (information) entropy}, defined as}
\begin{equation}
	H(X)=-\sum_{x}p(x)\log{p(x)},
\end{equation} 
where by convention, we use ``$\log$" to represent ``$\log_2$".
In general, $H(X)$ approximates the minimal binary description length $L$ of the random variable $X$, with the following inequality~\cite{Cover2006}: 
\begin{equation}
	H(X)\leq{L}<H(X)+1.
\end{equation}
It follows that, among all random variables with $c$ elements, the one with uniform distribution yields the maximum entropy, $\log(c)$.

Consider now two random variables $X$ and $Y$ with joint distribution 
\begin{equation}
	p(x,y)=\mbox{Prob}(X=x,Y=y),
\end{equation}
and conditional distribution 
\begin{equation}
	p(x|y)=\mbox{Prob}(X=x|Y=y). 
\end{equation}	
The {\it joint entropy} $H(X,Y)$ and {\it conditional entropy} $H(X|Y)$ for $X$ and $Y$ are defined, respectively, as
\begin{equation}
	H(X,Y) = -\sum_{x,y}p(x,y)\log{p(x,y)},
\end{equation}
and
\begin{equation}
	H(X|Y) = -\sum_{y}p(y)H(Y|X=x) = -\sum_{x,y}p(x,y)\log{p(x|y)}.
\end{equation}
Similar definition holds for $H(Y|X)$.

It is easy to verify that conditioning reduces entropy, i.e., knowledge of $Y$ will reduce (or at least cannot increase) the uncertainty about $X$, i.e.,  
\begin{equation}
	H(X|Y)\leq{H(X)}. 
\end{equation}	
Similarly, $H(Y|X)\leq{H(Y)}$.

The reduction of uncertainty of $X$ ($Y$) given full information about $Y$ ($X$) can be measured by the {\it mutual information} between $X$ and $Y$, as~\cite{Cover2006}
\begin{equation}
	I(X;Y) = H(X) - H(X|Y) = H(Y) - H(Y|X).
\end{equation}
The mutual information is symmetric in $X$ and $Y$, and measures their deviation from independence: if $X$ and $Y$ are fully dependent, then $H(X|Y)=H(Y|X)=0$ and thus $I(X;Y)=H(X)=H(Y)$; on the other hand, if $X$ and $Y$ are independent, then $H(X|Y)=H(X)$ and $H(Y|X)=H(Y)$ and therefore $I(X;Y)=0$.
In general, we have~\cite{Cover2006}
\begin{equation}
	0\leq I(X;Y)\leq\min[H(X),H(Y)].
\end{equation}

\begin{figure}[htbp]
\centering
\includegraphics*[width=0.8\textwidth]{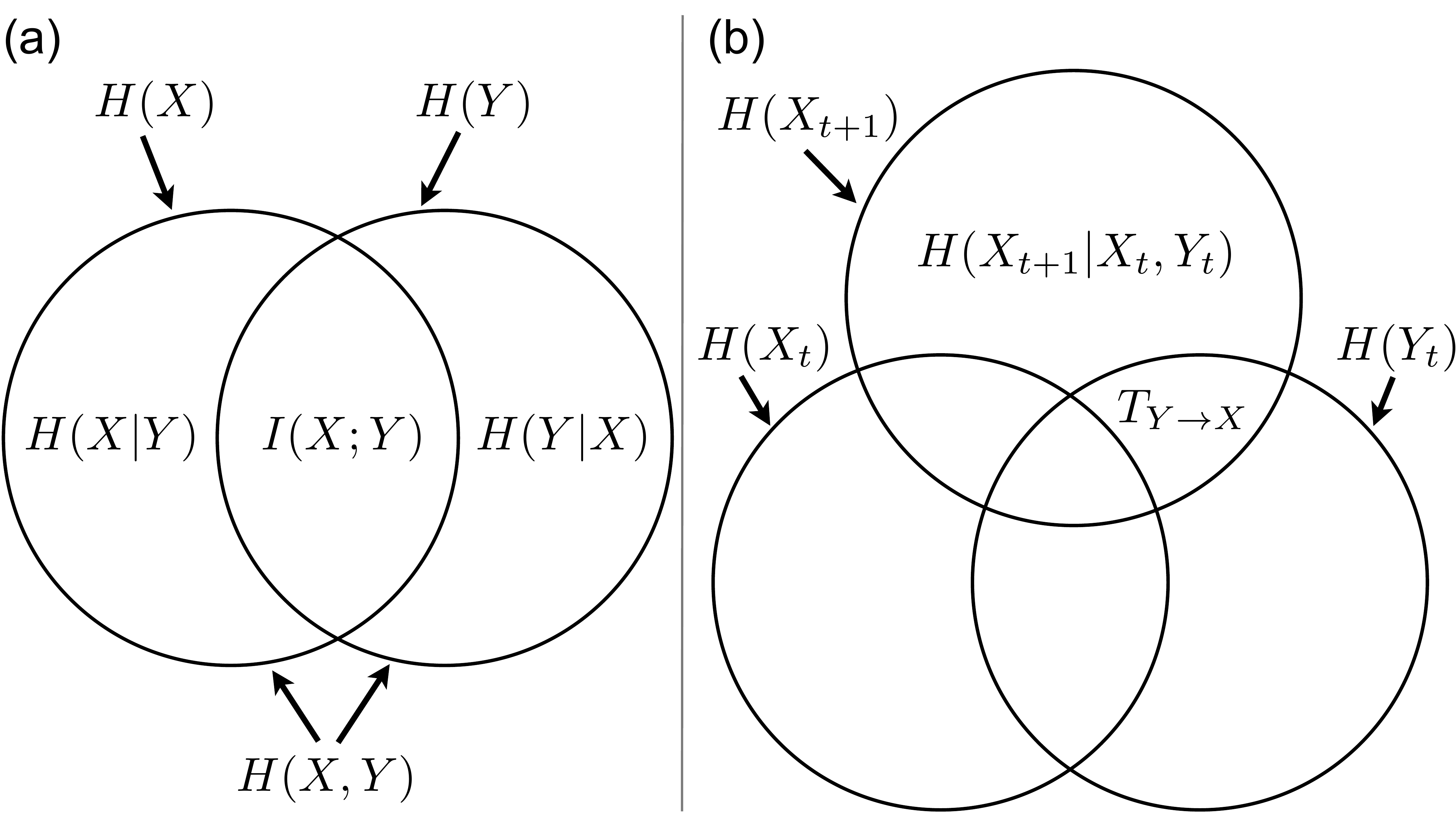}
\caption{Venn-like diagrams for information-theoretical measures.
(a) Relations between: entropies $H(X)$ and $H(Y)$, joint entropy $H(X,Y)$, conditional entropies $H(X|Y)$ and $H(Y|X)$, and mutual information $I(X;Y)$, of two random variables $X$ and $Y$.
(b) Relations between: transfer entropy $T_{Y\rightarrow{X}}$, entropies of random variables $X_{t+1}$, $X_t$, and $Y_t$, and their joint and conditional entropies. The transfer entropy is the difference between the conditional entropies $H(X_{t+1}|X_t,Y_t)$ and $H(X_{t+1}|X_t)$, which measures the extra information provided by $Y_t$ (in addition to $X_t$) in the determination of $X_{t+1}$.
}\label{fig:1}
\end{figure}

It is convenient to visualize the relationship between entropy, joint entropy, conditional entropy, and mutual information by a Venn-like diagram, as shown in Fig.~\ref{fig:1}(a).

We now turn to stochastic processes. For a stationary process $\{X_t\}$, its {\it entropy rate} $H(\mathcal{X})$ can be defined as
\begin{equation}
	H(\mathcal{X}) = \lim_{t\rightarrow\infty}H(X_t|X_{t-1},X_{t-2},\dots,X_1),
\end{equation}
which can be thought of as the (asymptotic) growth rate of the joint entropy $H(X_1,X_2,\dots,X_t)$.
If the process is Markovian, then~\cite{Cover2006}
\begin{equation}
	H(\mathcal{X})=\lim_{t\rightarrow\infty}H(X_t|X_{t-1}).
\end{equation}

For two stochastic processes $\{X_t\}$ and $\{Y_t\}$, the reduction of uncertainty about $X_{t+1}$ due to the information of
{the past $\tau_Y$ states of $Y$, represented by}
\begin{equation}
	Y^{(\tau_Y)}_t=(Y_t,Y_{t-1},\dots,Y_{t-\tau_Y+1}), 
\end{equation}
in addition to the information of {the past $\tau_X$ states of $X$, represented by}
\begin{equation}
	X^{(\tau_X)}_t=(X_t,X_{t-1},\dots,X_{t-\tau_X+1}), 
\end{equation}
is measured by the {\it transfer entropy} from $Y$ to $X$, defined as~\cite{Schreiber2000PRL}
\begin{equation}
	T_{Y\rightarrow X} = H(X_{t+1}|X^{(\tau_X)}_t) - H(X_{t+1}|X^{(\tau_X)}_t,Y^{(\tau_Y)}_t).\\
\end{equation}
One can similarly define $T_{X\rightarrow Y}$, which does not necessarily equal to $T_{Y\rightarrow X}$.
Note that $T_{Y\rightarrow X}$ can also be interpreted as the mutual information between $X_{t+1}$ and $Y^{(\tau_Y)}_t$ conditioned on $X^{(\tau_X)}_t$. In this paper, we focus on the case where 
\begin{equation}
	\tau_X=\tau_Y=1, 
\end{equation}
unless specified otherwise.
The relationship between transfer entropy, entropy and conditional entropy are illustrated in Fig.~\ref{fig:1}(b).

\section{Measuring Information Transfer in Two Coupled Oscillators}
\label{}
Coupled oscillator networks are commonly used for modeling the dynamic behavior of complex systems in various areas~\cite{Peroca1998PRL,Pikovsky2003,Arenas2008,Sun2009EPL}.
Here we consider discrete dynamics of coupled oscillator networks, in the form
\begin{equation}\label{eq:main}
	x^{(i)}_{t+1} = f[x^{(i)}_t] + \epsilon\sum_{j\neq{i}}c_{ij}g[x^{(i)}_t,x^{(j)}_t],~~i=1,2,\dots,N.
\end{equation}
Here $x^{(i)}_t\in\mathcal{D}\subset\mathbb{R}^d$ is the state of oscillator $i$ at time $t$, 
$f:\mathcal{D}\rightarrow\mathcal{D}$
is the dynamics of individual oscillators, 
$g:\mathcal{D}\times\mathcal{D}\rightarrow\mathcal{D}$
is the coupling function, and $\epsilon$ is the coupling strength.
Term $c_{ij}$ represents the coupling from $j$ to $i$.
In this paper, we use 
\begin{equation}
	f(x)=ax(1-x)
\end{equation}
with parameter $a=4$. The coupling function is chosen to be 
\begin{equation}
	g(x,y)=f(y)-f(x). 
\end{equation}
The choice of $\epsilon\in[0,1]$ and normalization condition 
\begin{equation}
	\sum_{j\neq{i}}c_{ij}=1
\end{equation}
guarantees that 
\begin{equation}
	x^{(i)}_t\in\mathcal{D}=[0,1]\mbox{~for all $i$ and $t$}.
\end{equation}

\begin{figure}[htbp]
\centering
\includegraphics*[width=0.8\textwidth]{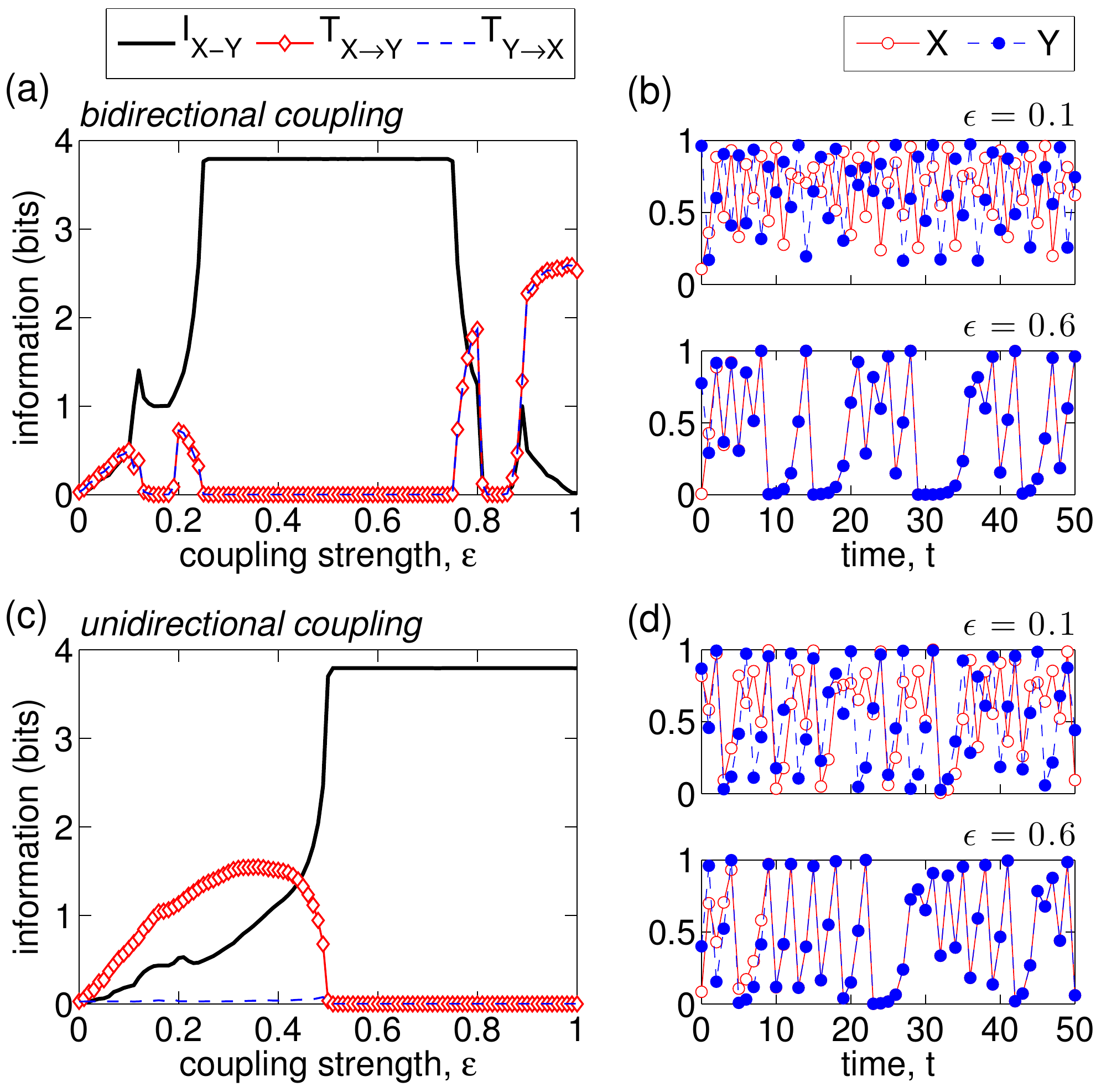}
\caption{Measuring information flow in two coupled logistic maps.
(a) Dependence of mutual information $I_{X-Y}=I(X;Y)$ and transfer entropies $T_{X\rightarrow{Y}}$ and $T_{Y\rightarrow{X}}$ on coupling strength $\epsilon$ for two bidirectionally coupled oscillators. Synchronization occurs when $\epsilon\in(0.25,0.75)$, which is the same region where the mutual information reaches its maximum. Due to the symmetry of coupling, $T_{X\rightarrow{Y}}=T_{Y\rightarrow{X}}$.
(b) Typical time series for two bidirectionally coupled oscillators, with $\epsilon=0.1$ (top, unsynchronized trajectories) and $\epsilon=0.6$ (bottom, synchronized trajectories).
(c-d) Same as (a-b), but for two oscillators with unidirectional coupling from oscillator $X$ to oscillator $Y$.
In this case, synchronization appears when $\epsilon\in[0.5,1)$. For $\epsilon\in(0,0.5)$, $T_{X\rightarrow{Y}}\gg T_{Y\rightarrow{X}}$, indicating that the dominant direction of information flow between $X$ and $Y$ is from $X$ to $Y$. 
In all simulations of the paper, we generate trajectories of length $10^5$ and discard the initial $5\%$ segments for all information measures. The interval $[0,1]$ is divided evenly into $2^4$ subintervals for the estimation of discrete probabilities.
{In our simulations, we made the choice of $2^4$ based on the balance between the length of the time series and the number of variables in the joint distribution: too few subintervals will only reveal limited information about the true dynamics and on the other hand, too many of them will lead to statistical under-sampling~\cite{Bollt2000PRL,Bollt2001PhysicaD}. Note that this problem of finding an appropriate number of subintervals for the estimation of entropy is analogous to the problem of finding an appropriate number of bins to construct a histogram, for which no ``best" solution exists in general.}
}\label{fig:2}
\end{figure}

We first explore information transfer in two coupled oscillators, with bidirectional and unidirectional couplings, respectively. With a slight abuse of notation, we use $X$ and $Y$ to represent oscillators $1$ and $2$. In terms of Eq.~\eqref{eq:main}, the bidirectional coupling corresponds to having $c_{12}=c_{21}=1$ and unidirectional coupling corresponds to $c_{12}=1, c_{21}=0$.
Results from numerical simulation are shown in Fig.~\ref{fig:2}.

One direct observation is that mutual information can be used as a measure of synchrony between two oscillators $X$ and $Y$. When $X$ and $Y$ are synchronized, their mutual information 
\begin{equation}
	I(X;Y)=H(X)=H(Y).
\end{equation}
When they are not synchronized, 
\begin{equation}
	I(X;Y)<\min[H(X),H(Y)].
\end{equation}
We remark that this observation suggests a new and alternative way of measuring generalized synchronization or  synchronization among a partial set of nodes in a large-scale network~\cite{Sun2009SIADS,Abrams2004}. 

For bidirectionally coupled oscillators, synchrony occurs when the coupling strength~\cite{2008JuangSIADS}
\begin{equation}
	\epsilon\in(0.25,0.75),
\end{equation}
as shown in Fig.~\ref{fig:2}(a).
The mutual information reaches its maximum for the same range of $\epsilon$.
Similarly, synchronization and maximum mutual information both occur when
\begin{equation}
	\epsilon\in(0.5,1],
\end{equation} 
in the case where $X$ and $Y$ are unidirectionally coupled [Fig.~\ref{fig:2}(c)].
Figure~\ref{fig:2}(b,d) show typical time series of the bidirectionally and unidirectionally coupled oscillators with $\epsilon=0.1$ (unsynchronized trajectories) and $\epsilon=0.6$ (synchronized trajectories), respectively.

When two oscillators synchronize, the transfer entropy from either one of them to the other becomes zero because no extra information can be gained by learning the past trajectory of the other oscillator (in addition to that from one's own). As a consequence, detection of coupling by transfer entropy (or any other measure) is valid only when the oscillators are not synchronized. Oscillators that are synchronized produce identical trajectories and therefore appear indistinguishable.

When the two oscillators are not synchronized, there is a positive transfer entropy following the directionality of coupling. For bidirectionally coupled oscillators,
\begin{equation}
	T_{X\rightarrow{Y}}=T_{Y\rightarrow{X}}>0\mbox{~if~}\epsilon\in(0,0.25)\cup(0.75,1],
\end{equation}
except for a few parameters at which the trajectories of $X$ and $Y$ settle into a periodic orbit [Fig.~\ref{fig:2}(a)]. For unidirectionally coupled oscillators, positive transfer entropy $T_{X\rightarrow{Y}}$ is observed when 
\begin{equation}
T_{X\rightarrow{Y}}>T_{Y\rightarrow{X}}=0\mbox{~if~}\epsilon\in(0,0.5).
\end{equation}
This absolute asymmetry of transfer entropy confirms the dominant direction of information flow from $X$ to $Y$, and not the other way around [Fig.~\ref{fig:2}(c)].


\section{Measuring Information Transfer in Coupled Oscillator Networks}
Having studied the application of transfer entropy in systems of two coupled oscillators, we now turn to networks.
\subsection{Effect of Indirect Influence}
\begin{figure}[htbp]
\centering
\includegraphics*[width=0.8\textwidth]{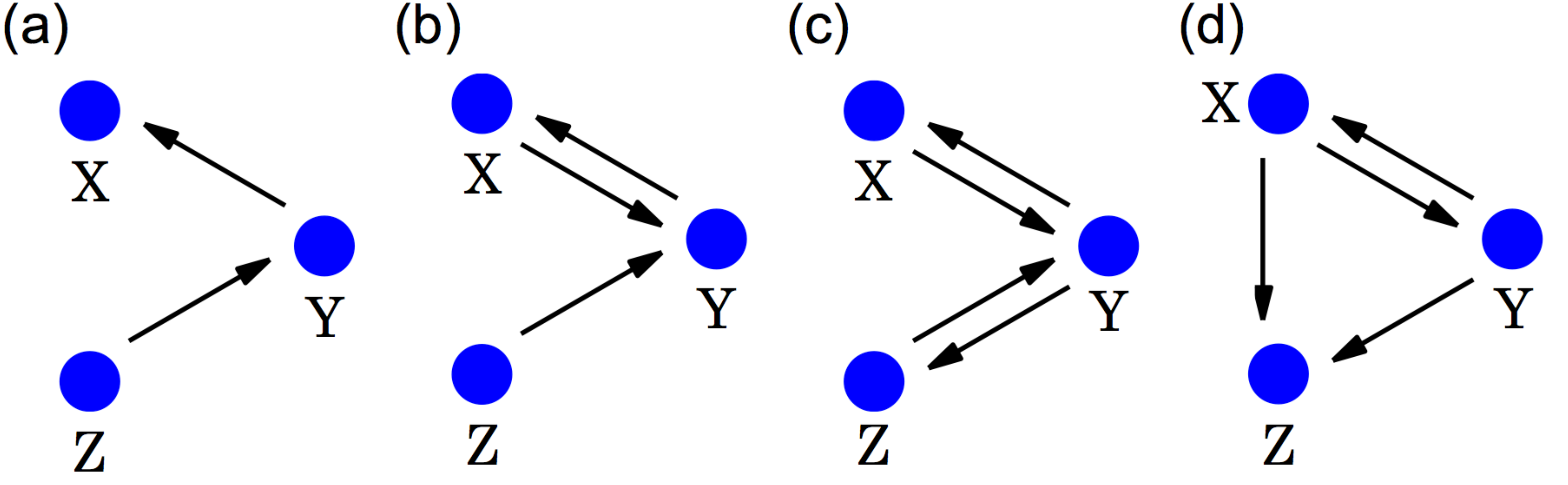}
\caption{Small-scale directed binary networks.
(a-c) Networks with a directed linear chain $Z\rightarrow{Y}\rightarrow{X}$, but no direct coupling $Z\rightarrow{X}$.
(d) A network that contains a direct coupling $X\rightarrow{Z}$, but not $Z\rightarrow{X}$.
}\label{fig:4}
\end{figure}

First we explore information transfer under the presence of indirect couplings. Consider a directed linear chain 
\begin{equation}
	Z\rightarrow{Y}\rightarrow{X},
\end{equation}
where $Z$ indirectly influences $X$ through $Y$ [Fig.~\ref{fig:4}(a)]. We focus on the dynamics of this three-node network according to Eq.~\eqref{eq:main}, with $\epsilon\in[0.2,0.4]$, a regime where coupling has a non-negligible effect on the dynamics but not strong enough to result in synchronization. 

In Fig.~\ref{fig:5}(a) we plot values of the transfer entropies $T_{X\rightarrow{X}}$, $T_{Y\rightarrow{X}}$, and $T_{Z\rightarrow{X}}$ ($T_{X\rightarrow{Y}}\approx T_{X\rightarrow{Z}}\approx 0$ are not plotted). 
By definition,  $T_{X\rightarrow{X}}=0$. The direct influence of $Y$ on $X$ is validated by the positive values of $T_{Y\rightarrow{X}}$.
Interestingly, values $T_{Z\rightarrow{X}}$ are also positive, despite the fact that there is no direct coupling from $Z$ to $X$.
Similar results are found for other networks that contain the direct linear chain $Z\rightarrow{Y}\rightarrow{X}$ but without the direct coupling 
$Z\rightarrow X$. See Fig.~\ref{fig:4}(b-c) for the other two networks and Fig.~\ref{fig:5}(c,e) for the corresponding results.

One important implication of these results is that, the use of transfer entropy for inferring network structure can be inappropriate under the presence of indirect influences. Since indirect couplings are common in many networks, 
{\it directed edges that are inferred by measuring transfer entropy can often be ``false positive".}

\subsection{Causation Entropy}
\begin{figure}[]
\centering
\includegraphics*[width=0.8\textwidth]{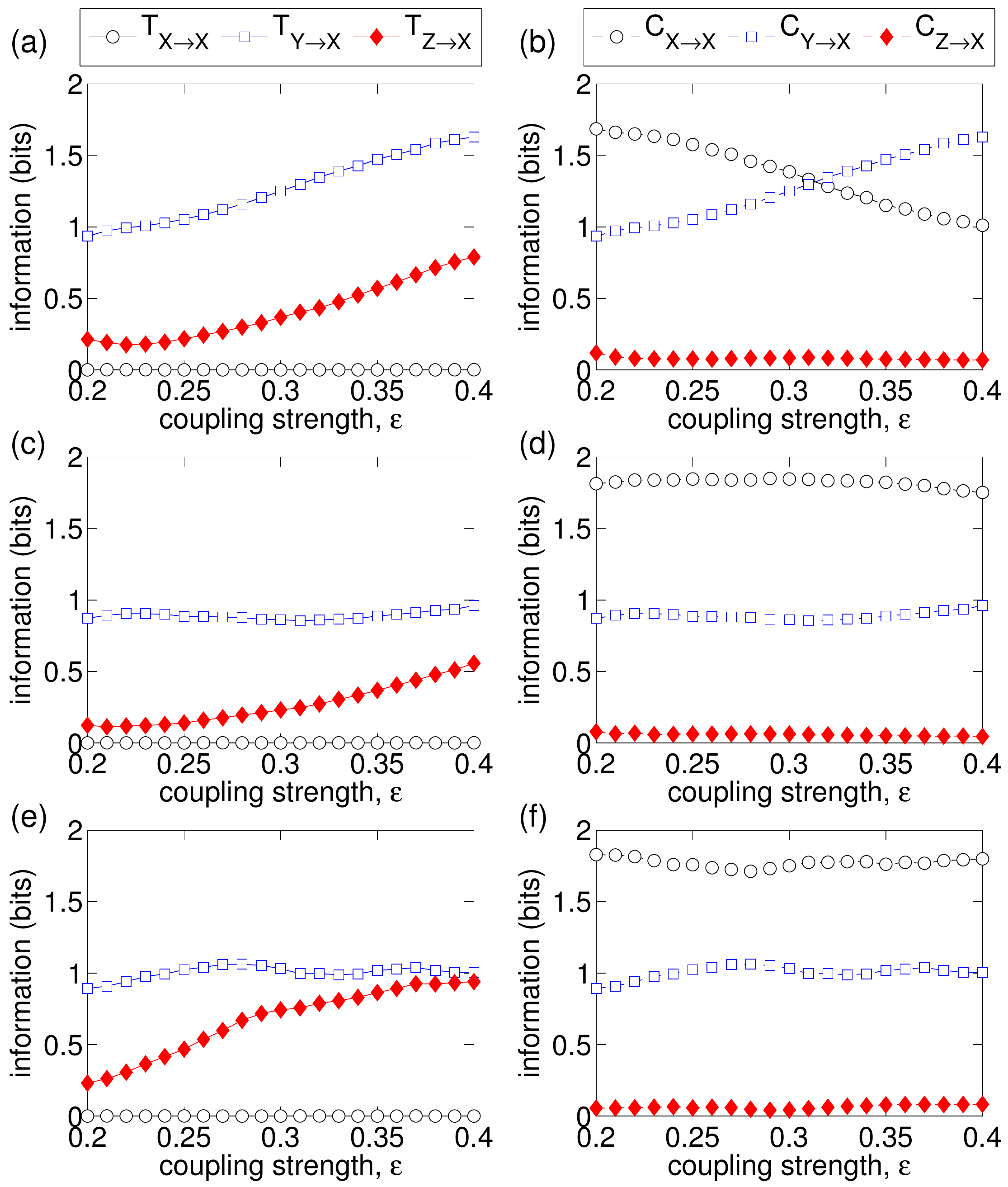}
\caption{Causation entropy and transfer entropy for the identification of indirect coupling.
(a-b) Transfer entropies $\{T_{X\rightarrow{X}},T_{Y\rightarrow{X}},T_{Z\rightarrow{X}}\}$ and
causation entropies $\{C_{X\rightarrow{X}},C_{Y\rightarrow{X}|(X)},C_{Z\rightarrow{X}|(X,Y)}\}$ for the network shown in Fig.~\ref{fig:4}(a) with dynamics~\eqref{eq:main}.
Note that the transfer entropy $T_{Z\rightarrow{X}}$ is positive despite the absence of direct coupling from $Z$ to $X$.
On the other hand, the causation entropy $C_{Z\rightarrow{X}|(X,Y)}\approx 0$, since information that are being indirected transferred from $Z$ to $X$ all go through $Y$.
(c-d) Same as (a-b), for the network in Fig.~\ref{fig:4}(b).
(e-f) Same as (a-b), for the network in Fig.~\ref{fig:4}(c).
}\label{fig:5}
\end{figure}

We note that the key reason transfer entropy often fails in identifying indirect couplings from direct ones is that it is a pairwise measure between two processes. For example, the transfer entropy $T_{Z\rightarrow X}$ shown in Fig.~\ref{fig:5}(a,c,e) does not account for the fact that the observed information transfer from $Z$ to $X$ is indeed a consequence of the direct information transfer from $Z$ to $Y$, and then $Y$ to $X$. 

Here we propose a new measure, which we call {\it causation entropy}.
The causation entropy from $Z$ to $X$ (conditioned on $X$ and $Y$) is defined as
\begin{equation}
	C_{Z\rightarrow X|(X,Y)} = H(X_{t+1}|X_t,Y_t) - H(X_{t+1}|X_t,Y_t,Z_t).
\end{equation}
Thus, $C_{Z\rightarrow X|(X,Y)}$ measures the extra information provided to $X$ by $Z$ {\it in addition} to the information that is already provided to $X$ by other means.

For an arbitrary set of processes, causation entropy is defined as follows.
\begin{mydef}[Causation Entropy]
The causation entropy from process $\mathcal{Q}$ to process $\mathcal{P}$ conditioned on the set of processes $\mathcal{S}$ is defined as
\begin{equation}
	C_{\mathcal{Q}\rightarrow\mathcal{P}|(\mathcal{S})} 
	= H(\mathcal{P}_{t+1}|\mathcal{S}_t) - H(\mathcal{P}_{t+1}|\mathcal{S}_t,\mathcal{Q}_t).
\end{equation}
\end{mydef}

Causation entropy $C_{\mathcal{Q}\rightarrow\mathcal{P}|(\mathcal{S})}$ is a generalization of transfer entropy. In fact, by letting $\mathcal{S}=\mathcal{P}$, we have
\begin{equation}
	C_{\mathcal{Q}\rightarrow\mathcal{P}|(\mathcal{P})}=T_{\mathcal{Q}\rightarrow\mathcal{P}}.
\end{equation}
In general, causation entropy $C_{\mathcal{Q}\rightarrow\mathcal{P}|(\mathcal{S})}$
measures the reduction in uncertainty in $\mathcal{P}$ due to the extra knowledge of $\mathcal{Q}$ in addition to that of $\mathcal{S}$. 

If $\mathcal{S}=\varnothing$,
we simply write 
\begin{equation}
	C_{\mathcal{Q}\rightarrow\mathcal{P}} = C_{\mathcal{Q}\rightarrow\mathcal{P}|(\varnothing)}.
\end{equation}
It follows that
\begin{equation}
	C_{\mathcal{Q}\rightarrow\mathcal{P}}= H(\mathcal{P}_{t+1}) - H(\mathcal{P}_{t+1}|\mathcal{Q}_t) =I(\mathcal{P}_{t+1};\mathcal{Q}_t),
\end{equation}
which is the mutual information between $\mathcal{P}_{t+1}$ and $\mathcal{Q}_t$.
When $\mathcal{S}\neq\varnothing$, causation entropy $C_{\mathcal{Q}\rightarrow\mathcal{P}|(\mathcal{S})}$ can be interpreted as the mutual information shared between $\mathcal{P}_{t+1}$ and $\mathcal{Q}_t$ conditioned on $\mathcal{S}_t$.

Figure~\ref{fig:5}(b,d,f) shows that, for the networks in Fig.~\ref{fig:4}(a-c), both $C_{X\rightarrow X}$ and $C_{Y\rightarrow X|(X)}$ are positive, as a result of the influence of $X$ on itself (self-dynamics) and the direct influence of $Y$ on $X$. On the other hand, and by design, the causation entropy $C_{Z\rightarrow X|(X,Y)}\approx 0$, in sharp contrast to the positive transfer entropy, $T_{Z\rightarrow X}>0$ [Fig.~\ref{fig:5}(a,c,e)].
The reason $C_{Z\rightarrow X|(X,Y)}$ is close to zero is that, the information provided by $Z$ (to $X$) is merely a subset of the information provided by $Y$. No extra information about $X$'s future state can be gained by learning the current state of $Z$ if those of $X$ and $Y$ are already known.

\subsection{Example: Dominance of neighbors}
Dominance of neighbors refers to a scenario where an oscillator's future state is dominantly determined by the state of its neighboring nodes, rather than by itself. In terms of Eq.~\eqref{eq:main}, this occurs when the coupling strength $\epsilon\approx 1$. We here explore its effect on information transfer. 
As an example, we consider dynamics by Eq.~\eqref{eq:main} on the network shown in Fig.~\ref{fig:4}(d), where node $X$ receives input from $Y$, but not from $Z$ (even indirectly).

\begin{figure}[htbp]
\centering
\includegraphics*[width=0.8\textwidth]{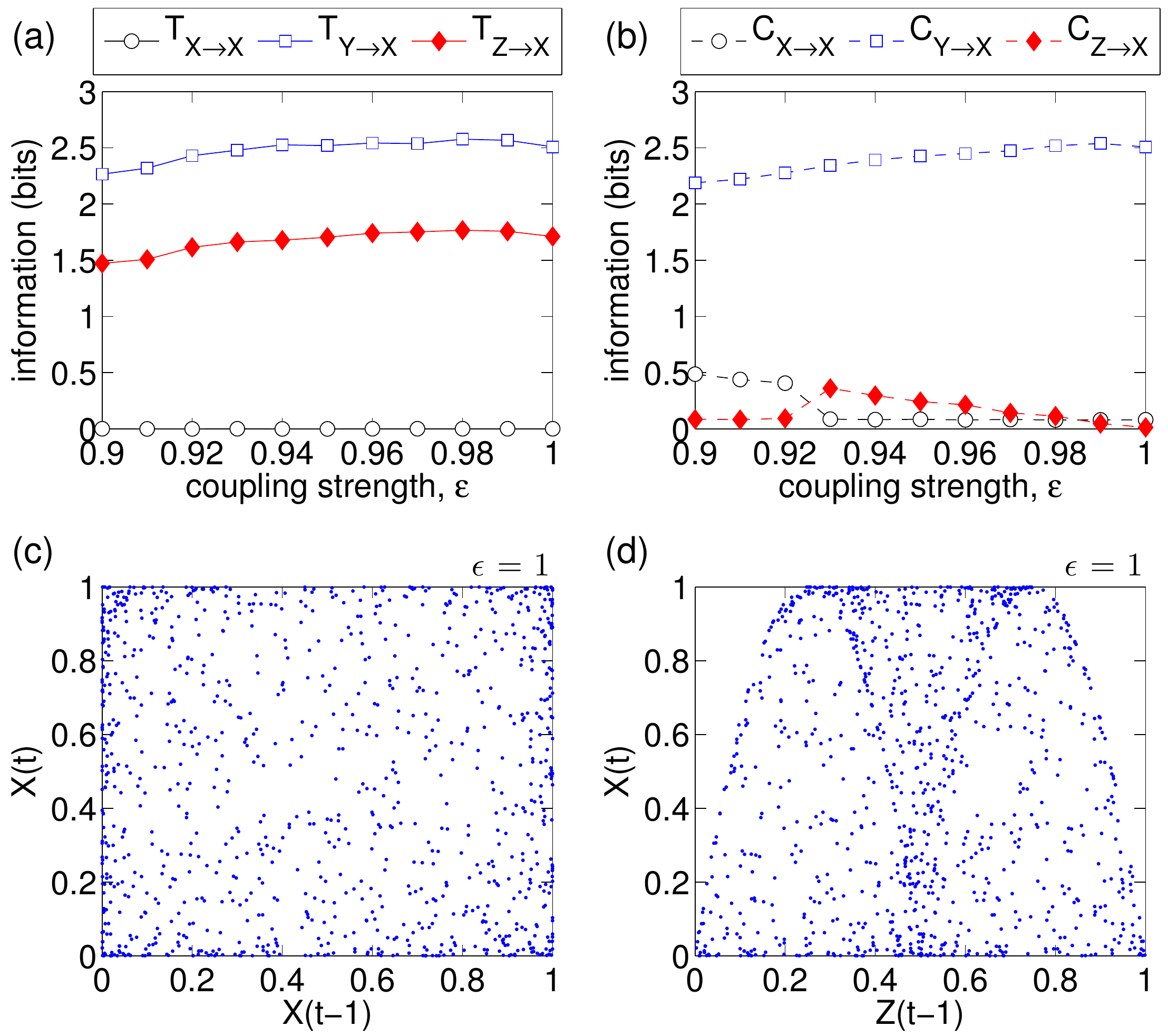}
\caption{Causation entropy versus transfer entropy under the dominance of neighbors.
(a) Transfer entropies $\{T_{X\rightarrow{X}},T_{Y\rightarrow{X}},T_{Z\rightarrow{X}}\}$.
(b) Causation entropies $\{C_{X\rightarrow{X}|(Y)},C_{Y\rightarrow{Y}},C_{Z\rightarrow{X}|(X,Y)}\}$ for the network in Fig.~\ref{fig:4}(d) whose dynamics follow Eq.~\eqref{eq:main}.
(c) Scatter plot between $X_{t-1}$ and $X_t$ for $\epsilon=1$.
(d) Scatter plot between $Z_{t-1}$ and $X_t$ for $\epsilon=1$.
In (c) and (d), points are taken from a randomly select trajectory segment (of the full trajectory) of length $1000$.
}\label{fig:6}
\end{figure}

As shown in Fig.~\ref{fig:6}(a), transfer entropy $T_{Y\rightarrow X}$ is positive, due to the direct influence of $Y$ on $X$. 
Surprisingly, transfer entropy $T_{Z\rightarrow X}$ is also found to be positive, despite the fact that no information flows from $Z$ to $X$, either directly or indirectly.

The reason positive transfer entropy $T_{Z\rightarrow X}$ is found in the absence of influence of $Z$ on $X$ is that, $T_{Z\rightarrow X}$ is taken to be the difference between $H(X_{t+1}|X_t)$ and $H(X_{t+1}|X_t,Z_t)$. Here since $X_{t+1}$ is dominantly determined by $Y_t$ and only depends weakly on $X_t$, the conditional entropies 
\begin{equation}
	\begin{cases}
		H(X_{t+1}|X_t)\approx H(X_{t+1}),\\
		H(X_{t+1}|X_t,Z_t)\approx H(X_{t+1}|Z_t).
	\end{cases}
\end{equation}
{A closer inspection of the network reveals that, under the strong coupling regime where the dynamics of an oscillator depends dominantly on its neighbors dynamics, the state of $X_{t+1}$ depends mostly on $Y_t$ (and not $X_t$). Since $Y_t$ depends mostly on $X_{t-1}$ by the very same argument, we conclude that the mutual information between $X_{t+1}$ and $X_{t-1}$ is high. Similarly, since $Z_t$ depends mostly on $X_{t-1}$ and $Y_{t-1}$, there is high mutual information between $Z_t$ and $X_{t-1}$. Based on this analysis, the mutual information between $X_{t-1}$ and $X_t$ should be low and that between $Z_t$ and $X_{t+1}$ should be nonnegilible, which is confirmed in Fig.~\ref{fig:6}(c-d). Although information in the network flows directly from $X$ to $Z$, without accounting for the dominant factors that determine the value of $X_{t+1}$, one would indeed infer a directed link from $Z$ to $X$ based on the calculation of the transfer entropy $T_{Z\rightarrow X}$.}

We note that, because of the dominance of $Y$ on $X$ (as opposed to $X$ on itself), one should indeed measure the causation entropies
$C_{Y\rightarrow X}$, $C_{X\rightarrow X|(Y)}$, and $C_{Z\rightarrow X|(X,Y)}$, respectively. Results are shown in Fig.~\ref{fig:6}(b).
The value $C_{Y\rightarrow X}>0$, as expected. The value $C_{X\rightarrow X|(Y)}\approx{0}$, due to the dominant influence of $Y$ (rather than $X$ itself) on $X$.
The value $C_{Z\rightarrow X|(X,Y)}\approx{0}$ as well, suggesting the absence of information transfer from $Z$ to $X$, which is consistent with the structure of the network shown in Fig.~\ref{fig:4}(d).

\subsection{Iterative Evaluation of Causation Entropy in a Network of $N$ Processes}
\label{}
The determination of causation entropies (i.e., the order $Y,X,Z$) can in fact be done {\it a priori}, by first choosing the process $\mathcal{Q}_1\in\{X,Y,Z\}$ that maximizes the causation entropy $C_{\mathcal{Q}_1\rightarrow X}$, and then iteratively select $\mathcal{Q}_k$ as the process that maximizes $C_{\mathcal{Q}_k\rightarrow X|(\mathcal{Q}_1,\dots,\mathcal{Q}_{k-1})}$ (see the following paragraph for details). For the example used in Fig.~\ref{fig:6}, we found that $\mathcal{Q}_1=Y$, and $\mathcal{Q}_2=X$. Therefore, contrast to transfer entropy, causation entropy can successfully identify the dominance of neighbors and in turn avoid erroneous inference of couplings due to its effect.

For a network of $N$ coupled stochastic processes $\{X^{(i)}_t\}_{i=1}^N$, we propose to identify the set of causal processes of a given process $i$ by iterative maximization of causation entropy. Let $n_0=i$. We first find process $n_1$ that satisfies
\begin{equation}
	n_1 = \operatorname{argmax}_{j\neq i}C_{X^{(j)}\rightarrow X^{(i)}}.
\end{equation}
Then we iteratively seek for $n_k$ ($k=2,3,\dots$) that satisfies
\begin{equation}
	n_k = \operatorname{argmax}_{j\neq i}C_{X^{(j)}\rightarrow X^{(i)}|(X^{(n_0)},X^{(n_1)},\dots,X^{(n_{k-1})})}.
\end{equation}
We stop the search at step $k$ when 
\begin{equation}
	C_{X^{(n_k)}\rightarrow X^{(i)}} < \theta,
\end{equation}
where $\theta$ is a preselected tolerance value.
The processes $n_1,n_2,\dots,n_{k-1}$ (in the decreasing order of dominance) form the set of causal processes of $i$.

{Note that in theory the value of $C_{X^{(n_k)}\rightarrow X^{(i)}}$ will be exactly zero if the dynamics of node $n_k$ does not causal-determine the dynamics of node $i$. In practice, however, the numerical estimation of $C_{X^{(n_k)}\rightarrow X^{(i)}}$ is based on the estimation of probability distributions from finite sample, and will be close to (but not necessarily equal to) zero for finite number of data points.
A rigorous way of determining whether the numerically computed causation entropy should be identified as zero is to perform a hypothesis test. It can be challenging to do such a test in practice and often times one can instead use a shuffle test to obtain approximate confidence intervals~\cite{Runge2012PRL}.}

\subsection{Example: Anticipatory couplings}
Our last example is a unidirectionally coupled dynamical system with anticipatory coupling~\cite{Hahs2011PRL}
\begin{equation}\label{eq:anti}
\begin{cases}
	x_{t+1} = f(x_t), \\
	y_{t+1} = (1-\epsilon)f(y_t) + \epsilon[(1-\alpha)f(x_t)+\alpha f^2(x_t)],
\end{cases}
\end{equation}
where $f(x)=ax(1-x)$ with $a=4$, parameter $\epsilon\in[0,1]$ is the coupling strength,
and parameter $\alpha\in[0,1]$ is the strength of anticipatory coupling. Notation $f^2$ means that the map $f$ is applied twice.

\begin{figure}[htbp]
\centering
\includegraphics*[width=0.8\textwidth]{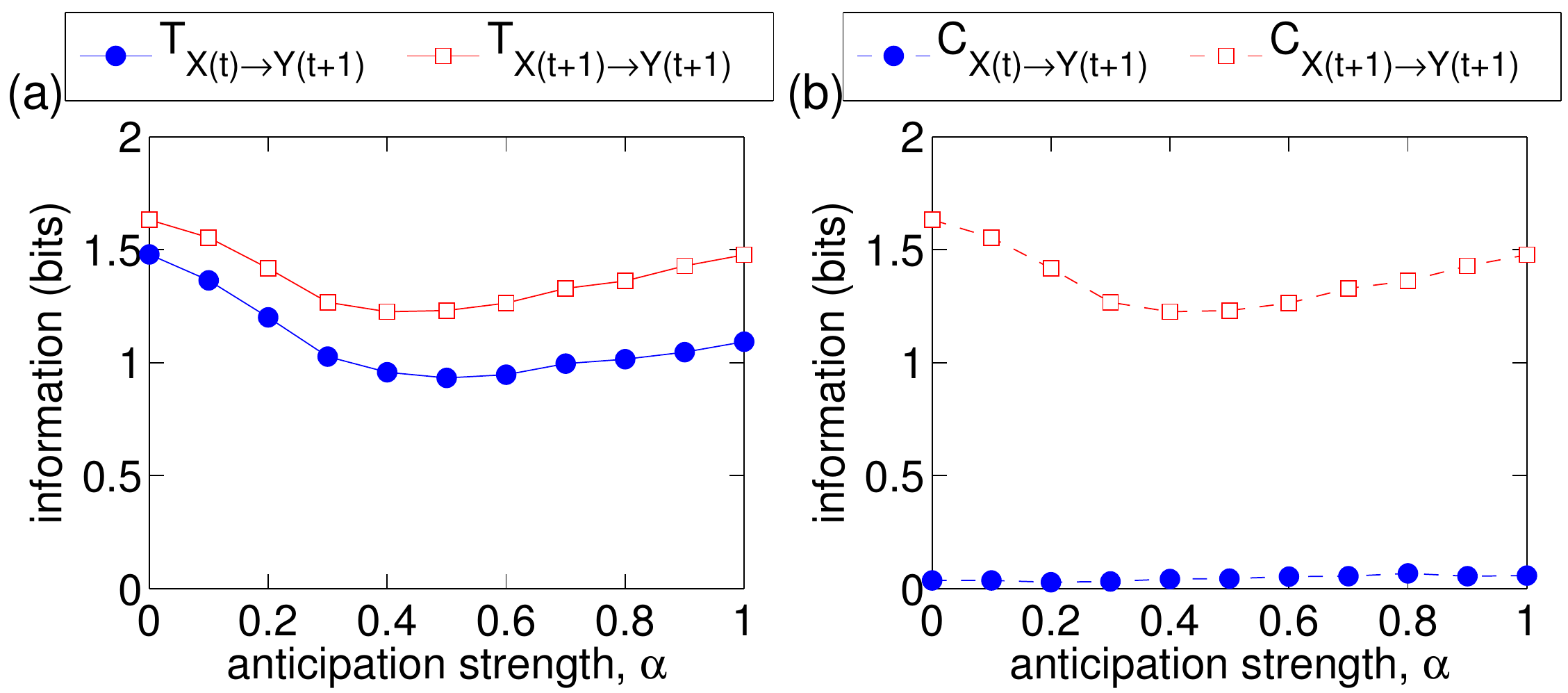}
\caption{Causation entropy versus transfer entropy under anticipatory coupling.
(a) Transfer entropies $T_{X_{t}\rightarrow Y_{t+1}}$ and $T_{X_{t+1}\rightarrow Y_{t+1}}$.
(b) Causation entropies $C_{X_{t+1}\rightarrow Y_{t+1}|(Y_t)}$ and $C_{X_{t}\rightarrow Y_{t+1}|(Y_t,X_{t+1})}$.
Here parameter $\epsilon=0.3$.
}\label{fig:7}
\end{figure}

Here we adopt the concept and notation of transfer entropy to define
\begin{equation}
	T_{X_{t}\rightarrow Y_{t+1}}=T_{X\rightarrow Y}=H(Y_{t+1}|Y_t)-H(Y_{t+1}|Y_t,X_t),
\end{equation}
and 
\begin{equation}
	T_{X_{t+1}\rightarrow Y_{t+1}}=H(Y_{t+1}|Y_t)-H(Y_{t+1}|Y_t,X_{t+1}).
\end{equation}
Figure~\ref{fig:7}(a) shows that both $T_{X_{t}\rightarrow Y_{t+1}}$ and $T_{X_{t+1}\rightarrow Y_{t+1}}$ are positive, with comparable values.
Does this suggest that both $X_t$ and $X_{t+1}$ independently influence $Y_{t+1}$?
Standard interpretation (of transfer entropy) would suggest that the answer to this question is yes.

By use of causation entropy, we find that $Y_{t+1}$ is primarily determined by $Y_t$. The second dominant influence on $Y_{t+1}$ is $X_{t+1}$, as confirmed by the values of 
\begin{equation}
	C_{X_{t+1}\rightarrow Y_{t+1}|(Y_t)}=H(Y_{t+1}|Y_t)-H(Y_{t+1}|Y_t,X_{t+1}).
\end{equation}
It turns out that additional information of $X_t$ (beyond $Y_t$ and $X_{t+1}$) does not contribute to the reduction of uncertainty of $Y_{t+1}$. This is validated by the causation entropy
\begin{equation}
	C_{X_{t}\rightarrow Y_{t+1}|(Y_t,X_{t+1})}=H(Y_{t+1}|Y_t,X_{t+1})-H(Y_{t+1}|Y_t,X_{t+1}),
\end{equation}
which remain close to zero, as shown in Fig.~\ref{fig:7}(b).

Therefore, in contrast to transfer entropy analysis, which would suggest that both $X_t$ and $X_{t+1}$ participate in the determination of $Y_{t+1}$, causation entropy analysis reveals that information of $X_t$ is indeed completely redundant in inferring $Y_{t+1}$. In fact, by expressing $f(x_t)$ as $x_{t+1}$ in Eq.~\eqref{eq:anti}, it appears that the value of  $y_{t+1}$ depends solely on $y_t$ and $x_{t+1}$, and not on $x_t$.

\section{Information Transfer in Time-Dependent Networks}
The effects of time-dependent structures on network dynamics are often intriguing and pose considerable challenges for analysis. For example, the problem of synchronization stability of coupled oscillators in time-dependent networks has been fully addressed only for a few specific cases~\cite{Belykh2004,Porfiri2006,Stilwell2006SIADS,Sorrentino2008,Taylor2010}. Here, our focus is to measure information transfer among oscillators that are coupled through a time-dependent network structure (that is, a network whose edges change in time). In particular, we generalize Eq.~\eqref{eq:main} to allow for time-dependent interactions in between oscillators, as
\begin{equation}\label{eq:main2}
	x^{(i)}_{t+1} = f[x^{(i)}_t] + \epsilon\sum_{j\neq{i}}c_{ij}(t)g[x^{(i)}_t,x^{(j)}_t],~~i=1,2,\dots,N.
\end{equation}
Here all terms in Eq.~\eqref{eq:main2} except for $c_{ij}(t)$ are the same as those in Eq.~\eqref{eq:main}. The term $c_{ij}(t)$ represents the coupling from $j$ to $i$ at time $t$ and explicitly accounts for the time-dependent network structure.

\begin{figure}[htbp]
\centering
\includegraphics*[width=0.8\textwidth]{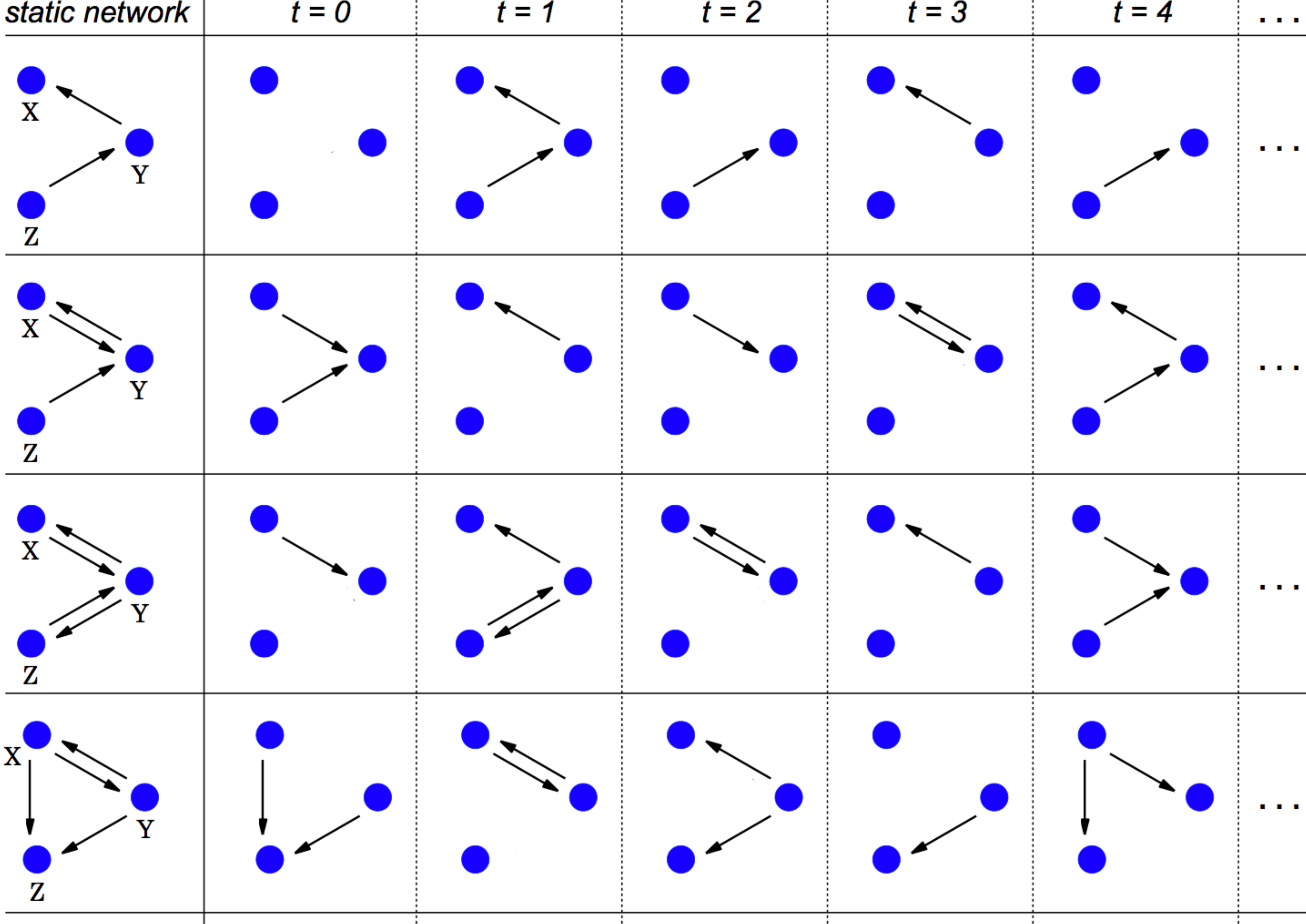}
\caption{Examples of time-dependent networks. First (leftmost) column: structure of static networks (the same as those in Fig.~\ref{fig:4}). Second to the last (rightmost) columns: typical network structures at different times, obtained from keeping each directed edge of the static network independently with probability $p=0.5$ at each time $t$.
}\label{fig:9}
\end{figure}

We consider time-dependent networks constructed as follows. Start with a baseline static network whose adjacency matrix is $\bar{C}=[\bar{c}_{ij}]_{n\times n}$. The edges in the network are then allowed to ``blink" according to the following rule, to generate a time-dependent network: at each time $t$,  
\begin{equation}\label{eq:blinking}
c_{ij}(t) = 
\begin{cases}
	\bar{c}_{ij}, & \mbox{with probability $p$;}\\
	0, & \mbox{with probability $1-p$.}
\end{cases}
\end{equation}
Therefore, when the blinking probability $p=1$, the network is the same as the baseline static network; on the other extreme, when $p=0$, no edge exists and the network becomes empty (i.e., each oscillator is isolated and does not couple to other oscillators). For the values of $p$ in between $0$ and $1$, the network structure changes in time in a stochastic fashion (see Fig.~\ref{fig:9} for a few illustrative examples).

Our interest lie in the information transfer within such time-dependent networks. Different from its static counterpart, the flow of information in a time-dependent network often cannot be directly obtained from examining the edges $c_{ij}(t)$, because it is possible for a network to be disconnected at all times and yet be able to transfer information from one node to another. Such scenario has been previously considered in the synchronization of coupled oscillators in time-dependent networks with edges being switched on and off~\cite{Stilwell2006SIADS} and in moving-neighbor networks whose edges are defined by the local interactions between agents that move in space~\cite{Porfiri2006}. In both cases, even though the original static network is connected, the corresponding time-dependent network obtained by blinking the edges might not be (see Fig.~\ref{fig:9} for examples). 

The connection between these time-dependent networks and the original static network is that the asymptotic temporal average of each directed edge, $\langle c_{ij}\rangle$, is proportional to the weight of the same edge in the static network:
\begin{equation}
	\langle c_{ij}\rangle = \lim_{T\rightarrow\infty}\frac{1}{T}\sum_{t=1}^{T}c_{ij}(t) = p\bar{c}_{ij}.
\end{equation}

\begin{figure}[htbp]
\centering
\includegraphics*[width=0.8\textwidth]{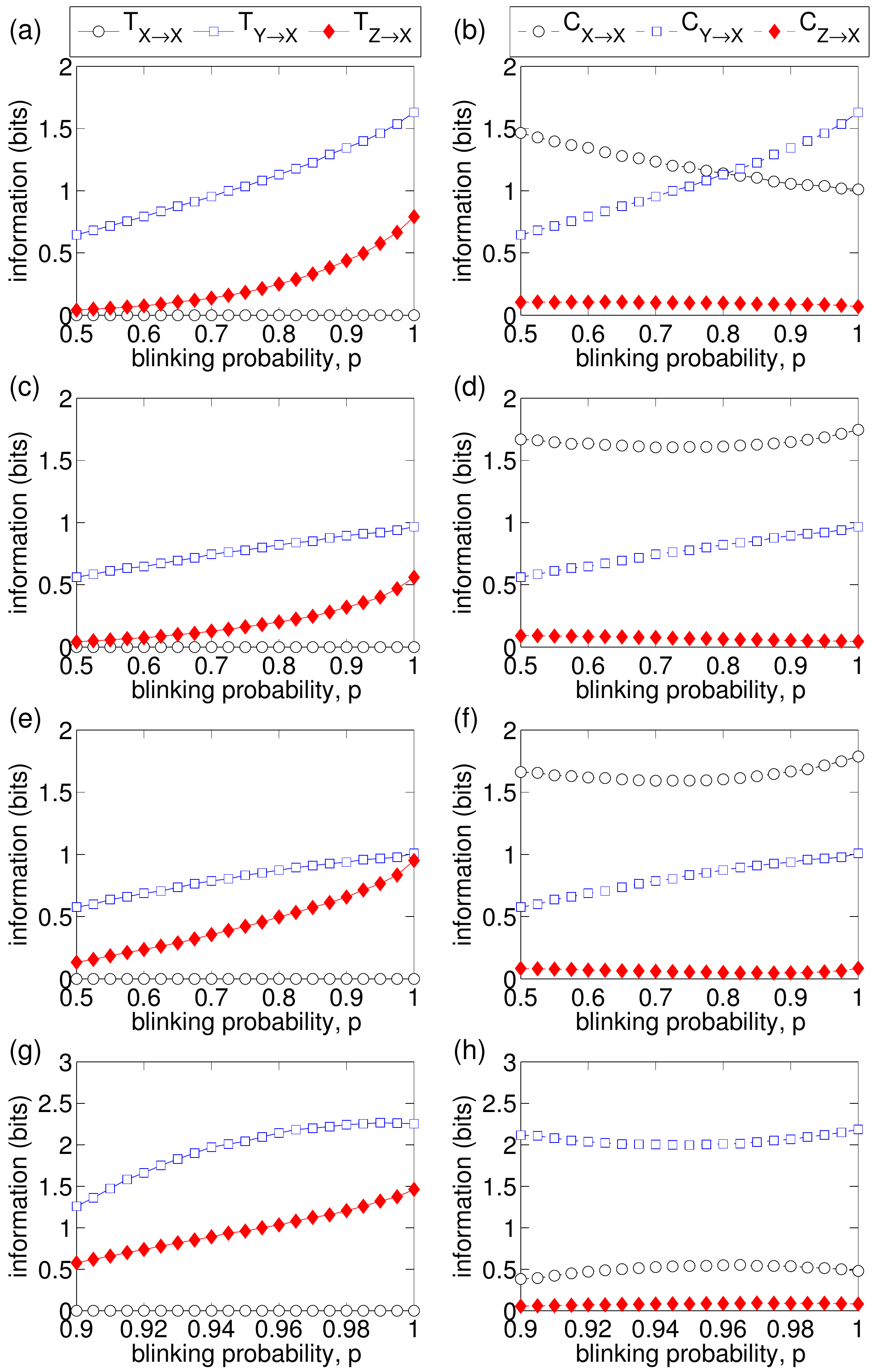}\vspace{-0.1in}
\caption{Causation entropy versus transfer entropy for time-dependent networks.
(a-b) Transfer entropies $\{T_{X\rightarrow{X}},T_{Y\rightarrow{X}},T_{Z\rightarrow{X}}\}$ and
causation entropies $\{C_{X\rightarrow{X}},C_{Y\rightarrow{X}|(X)},C_{Z\rightarrow{X}|(X,Y)}\}$ for the time-dependent network originates from the network  in Fig.~\ref{fig:4}(a) via Eq.~\eqref{eq:blinking} and endowed with dynamics~\eqref{eq:main2}, for the fixed coupling strength $\epsilon=0.4$.
(c-f) Same as (a-b), for the networks in Fig.~\ref{fig:4}(b) and Fig.~\ref{fig:4}(c), respectively, and $\epsilon=0.4$.
(g-h) Same as (a-b), for the network in Fig.~\ref{fig:4}(d) and $\epsilon=0.9$.
}\label{fig:8}
\end{figure}

We ran numerical simulation on several time-dependent networks and focus on the information flow measured both by transfer entropy and causation entropy. Figure~\ref{fig:8}(a) shows that, for the directed linear chain $Z\rightarrow{Y}\rightarrow{X}$ with fixed coupling strength $\epsilon=0.4$, when the blinking probability $p$ increases, the transfer entropy $T_{Z\rightarrow{X}}$ becomes increasingly nonnegligible, indicating direct information transfer from $Z$ to $X$ from standard interpretation. On the other hand, the causation entropy $C_{Z\rightarrow{X}}$ remains essentially zero, suggesting that the information transferred from $Z$ to $X$ is merely a redundancy of the information that are transferred from $Z$ to $Y$ and $Y$ to $X$, respectively, possibly at different times. Figure~\ref{fig:8}(b-c) show similar comparison between transfer entropy and causation entropy for meaning the information flow in time-dependent networks that originate from the networks shown in Fig.~\ref{fig:4}(b-c) with the fixed coupling strength $\epsilon=0.4$.

The possible misinterpretation of transfer entropy becomes more evident under the dominance of neighbors scenario, where the coupling strength $\epsilon$ is close to $1$. As shown in Fig.~\ref{fig:8}(d), under such scenario, transfer entropy identifies a strong information transfer from $Z$ to $X$ whereas in the average network of the time-dependent network, it is the exact opposite. Causation entropy, on the other hand, successfully identifies the dominant nodes that influence the dynamics of $X$, namely, its neighbor $Y$ and then $X$ itself.



\section{Discussion and Conclusion}
Our main message here is that while being an essential problem in science in general, and dynamical systems in particular, the question of what is cause and what is influence in complex system analysis is challenging, not due to the lack of methodology, but rather due to the lack of clear and comprehensible understanding of the applicability of proposed methods, in particular when the underlying system involves complex interactions.
The popular concept of transfer entropy has been used lately to serve as a way of inferring causality, without much understanding about its domain of success.

We here explored information flow measured from the dynamics of small-scale coupled oscillators network, attempting to gain insights into the validity of transfer entropy as well as its limitations.
For two coupled oscillators, transfer entropy is found to successfully detect the directionality of information flow, even in cases where the couplings are blinking (time-dependent). However, its validity breaks down under the presence of indirect couplings, dominance of neighboring dynamics, and anticipatory couplings, which are common in large-scale complex systems.

To overcome the limitations of transfer entropy, we introduced a new measure of information flow called causation entropy, which is designed to allow inference of causation despite the presence of primary and secondary influences between elements of a larger coupled system.
We highlighted the success of our approach with several examples where specifically the transfer entropy cannot distinguish between causation and independence but causation entropy successfully infers the true causal relationships. 

Given the recent advancements in estimating transfer entropy in rather general settings including multivariate time series and infinite time delay~\cite{Runge2012PRL,Ahmed2011PRE}, it is our hope to build on the idea of causation entropy to explore information flow and coupling inference in larger-scale systems, which are important for a wide range of applications across scientific fields.
{One challenge is that, for large-scale systems, naive binning methods would require an exponential number of data points with respect to the number of variables, in order to reliably calculating entropies (including joint entropy, transfer entropy, and also causation entropy). Nonparametric density estimation methods previously developed for mutual information~\cite{Schindler2007,Kraskov2004PRE} are likely to offer a route towards the reliable estimation of causation entropies in large-scale dynamical systems.}

\section*{Acknowledgments}
We thank Dr Samuel Stanton from the ARO Complex Dynamics and Systems Program for his ongoing and continuous support. This work is funded by ARO Grant No. 61386-EG.



\newpage

\end{document}